\documentclass[a4paper, 10pt, 5p,preprint,num]{elsarticle}
\biboptions{sort&compress}
\usepackage[utf8]{inputenc}

\usepackage{amsmath}
\usepackage{amssymb}
\usepackage{bbm}
\usepackage{bm}
\usepackage{color}
\numberwithin{equation}{section}
\numberwithin{figure}{section}
\usepackage{physics}
\usepackage{graphicx}
\usepackage{epstopdf}
\usepackage{aligned-overset}
\usepackage{hyperref}
\hypersetup{colorlinks=true,linktoc=page,linkcolor=blue,citecolor=red,urlcolor=cyan}

\usepackage[english]{babel}

\newcommand{\hu}{\hspace{0.6cm}}

\newcommand{\dx}[1][x]{{{\rm d} #1}}
\newcommand{\mathi}{{\text{i}}}

\newcommand{\dime}{d}

\usepackage{xparse}

\NewDocumentCommand{\dkpara}{O{k} O{\dime}}{\frac{{\rm d} {#1}^{\parallel}}{(2\pi)^{#2}}}
\NewDocumentCommand{\dkd}{O{k} O{\dime}}{ \frac{{\rm d}^{#2} {#1}}{(2\pi)^{#2}}}
\NewDocumentCommand{\dxd}{O{x} O{\dime}}{ {\rm d}^{#2} {#1} }
\NewDocumentCommand{\ad}{O{} O{}}{ \bigl \langle {#1}\bigr\rangle_{{\rm ad}{#2}} }

\title{Resummed heat kernel and effective action for Yukawa and QED}
\author[1,2]{S.~A.~Franchino-Viñas\corref{cor1}%
}

\author[2,3]{C.~Garc\'ia-P\'erez}

\author[4,5]{F.~D.~Mazzitelli}

\author[2,3]{V.~Vitagliano}
\author[6,7]{U.~Wainstein-Haimovichi}

\cortext[cor1]{Corresponding author}

\affiliation[1]{organization={Helmholtz-Zentrum Dresden-Rossendorf},
addressline={Bautzner Landstra{\ss}e 400},
postcode={01328},
city={Dresden},
country={Germany}}

\affiliation[2]{organization={DIME, Università di Genova},
addressline={Via all'Opera Pia 15},
postcode={16145},
city={Genova},
country={Italy}}

\affiliation[3]{organization={INFN Sezione di Genova},
addressline={Via Dodecaneso 33},
postcode={16146},
city={Genova},
country={Italy}}

\affiliation[4]{organization={Centro Atómico Bariloche, CONICET, Comisión Nacional de Energñia Atómica},
addressline={Av. Bustillo Km. 9,5},
postcode={R8402AGP},
city={Bariloche},
country={Argentina}}

\affiliation[5]{organization={Instituto Balseiro, Universidad Nacional de Cuyo},
addressline={Av. Bustillo Km. 9,5},
postcode={R8402AGP},
city={Bariloche},
country={Argentina}}

\affiliation[6]{organization={Departamento de Física, Facultad de Ciencias Exactas, Universidad Nacional de La Plata},
addressline={C.C.\ 67},
postcode={1900},
city={La Plata},
country={Argentina}}

\affiliation[7]{
organization={Instituto de Física La Plata (UNLP-CONICET)},
addressline={Diagonal 113 e/63 y 64},
postcode={1900},
city={La Plata},
country={Argentina}}

\begin{document}

\begin{abstract}
In this letter, we prove the existence of resummed expressions for the diagonal of the heat kernel and the effective action of a quantum field which interacts with a scalar or an electromagnetic background. Working in an arbitrary number of spacetime dimensions, we propose an Ansatz beyond the Schwinger--DeWitt proposal, effectively resumming an infinite number of invariants which can be constructed from powers of the background, as well as its first and second derivatives in the Yukawa case.
This provides a proof of the recent conjecture that all terms containing the invariants
$F_{\mu\nu}F^{\mu\nu}$ and $\widetilde F_{\mu\nu}F^{\mu\nu}$ in the proper-time series expansion of the SQED effective action can be resummed.
Possible generalizations and several applications are also discussed---in particular, the existence of an analogue of the Schwinger effect for Yukawa couplings.

\end{abstract}


\maketitle


\section{Introduction}\label{sec:intro}

Understanding nonperturbative phenomena stands out as a paramount challenge in contemporary high-energy physics. While the perturbative approach to quantum field theory has become a standard tool, efficient nonperturbative techniques are still under active development.
Systematic theoretical methods such as the functional renormalization group approach~\cite{Bonanno:2020bil}, the Dyson--Schwinger equations~\cite{Fischer:2018sdj} or simply lattice numerical techniques~\cite{ExtendedTwistedMass:2022jpw} are currently employed to gain valuable insights into the intricate realm of the quantum world.

Investigating a single quantum field interacting with background classical fields appears perhaps as a relatively simpler problem. However, even in this case, the available toolkit is relatively limited, often resorting to methods like the WKB approach~\cite{Dunne:1998ni}, the computation of instantons~\cite{Dunne:2006st,Schutzhold:2008pz}, etc.
In this semiclassical scenario, the classical fields are assumed fixed by sources that can be experimentally tuned at will. Notably, certain regimes provide an ideal testbed to probe quantum effects, as seen, for instance, in the context of QED in the presence of intense background fields~\cite{Fedotov:2022ely}, or in settings involving static~\cite{Bordag:2009zz} and dynamical Casimir effects~\cite{Dalvit:2010ria, Nation:2011dka,Dodonov:2020eto}. Another significant example is the case of gravitational background fields, with applications to cosmology, astrophysics and black hole evaporation~\cite{Birrel:1982, Parker:2009}.

Within the framework of single quantum fields interacting with classical backgrounds, we will explore a scenario involving a quantum, complex scalar field $\phi$ minimally coupled to an electromagnetic field $\mathcal{A}_\mu$ and interacting with a scalar field $H$ through a Yukawa term,
\begin{align} S:= \frac{1}{2}\int \dxd[x][\dime] \Big[ \vert (\nabla-\mathcal{A}(x)) \phi(x) \vert^2+H^2(x) \vert \phi(x) \vert^2 \Big].\end{align}
Here, $X^2 := X_\mu X^\mu$ and we assume $\mathcal{A}$ to be anti-selfadjoint. While the fields $H$ and $\mathcal{A}_\mu$ could in principle be quantized, our primary focus is on the semiclassical theory. Therefore we replace them with classical fields $\sqrt{V}$ and $A_\mu$ (their vacuum expectation values), assuming only their regularity as needed. A few further comments are in order. First, $V$ and $A_\mu$ are not going to be considered on-shell and Bianchi identities will be employed only to compare our results with the existing literature.
Additionally, $V$ may include a mass term for the quantum field $\phi$
(or, looking from the opposite perspective, if there were a mass term, it could be absorbed into $V$).
Finally, it's worth stressing that we will work in flat space, leaving potential generalizations to curved space for future discussions.

In addressing the current issue, standard quantization procedures can be applied. This leads us to the formulation of the one-loop effective action $\Gamma$, which can be straightforwardly connected to the quantum fluctuations operator $\mathcal{Q}$,
\begin{align}\label{eq:EA_logdet}
 \Gamma&= \operatorname{Log} \operatorname{Det} \left[\mathcal{Q}\right],
 \\
 \mathcal{Q}:&=-(\nabla - A )^2+V(x).\label{eq:op}
\end{align}
Using the Schwinger--DeWitt (SDW) techniques~\cite{DeWitt:2003}, we can recast the effective action in terms of the trace of the related heat kernel operator\footnote{This is an extension of Frullani's representation, which is valid for the logarithm of a quotient~\cite{Jeffreys}.
To be more precise, one could take the quotient with the quantum fluctuations operator corresponding to the background-free case.} $K$;
in a $\dime$-dimensional Euclidean space we obtain:
\begin{align}\label{eq:effective_action_hk}
\begin{split}
 \Gamma  &= -\int_0^{\infty} \frac{\dx[\tau]}{\tau} \int \dxd[x][d] K(x,x;\tau),
 \\
 K(x,x';\tau) &= e^{- \tau \mathcal{Q}}(x,x';\tau).
\end{split}
\end{align}
These expressions enable us to employ the full range of mathematical tools available in the study of the heat kernel and derive the consequences at the level of the effective action.

Our attention is particularly directed towards situations where the standard proper-time expansion of the heat kernel in terms of the Gilkey--Seeley--DeWitt (GSDW) coefficients turns out to be inadequate and
one is forced to recur to a {\em resummed} version, as a first approach to analyze various nonperturbative aspects of complicated quantum field theories. A few results have already been established in this direction. For instance, in the realm of the first quantization, the heat kernel can be understood as a propagator in imaginary time, for which a resummation in the potential holds~\cite{Wigner:1932eb}.
In quantum field theory, on the one hand, the covariant perturbation theory outlined in~\cite{Barvinsky:1987uw, Barvinsky:1990up} is applicable in cases involving rapidly oscillating curvatures or potentials and allows, for example, to obtain beta functions in a momentum-like scheme~\cite{Gorbar:2002pw, Gorbar:2003yt, Franchino-Vinas:2018gzr,Silva:2023lts}.
On the other hand, the resummation of the Ricci scalar conjectured in Ref.~\cite{Parker:1984dj} (and later proved in~\cite{Jack:1985mw}) has been used to establish several properties of fermionic condensates in curved space \cite{Flachi:2014jra, Flachi:2015sva, Flachi:2019btk, Castro:2018iqt}. Similar resummation techniques have also been employed in the investigation of Casimir self-interactions under spacetime-dependent boundary conditions \cite{Franchino-Vinas:2020okl, Edwards:2021cyp, Ahmadiniaz:2022bwy} and, more in general,
for background fields that are simple enough to obtain a closed expression either for the effective action~\cite{Heisenberg:1936nmg, Weisskopf:1936hya, Brown:1975bc} or for the corresponding heat kernel~\cite{Avramidi:1995ik, Avramidi:2009quh}.

More recently, it has been conjectured that for large field strengths, the fundamental  invariants $\mathcal F=F_{\mu\nu}F^{\mu\nu}$ and $\mathcal G=\widetilde{F}_{\mu\nu}F^{\mu\nu}$ appearing in the proper-time expansion of the (S)QED one-loop effective action can be resummed~\cite{Navarro-Salas:2020oew}, resulting in a local version of the Euler--Heisenberg prefactor in the effective Lagrangian, originally derived in Refs.~\cite{Heisenberg:1936nmg, Weisskopf:1936hya, Schwinger:1951nm}.
The first main result in this letter
is the statement and the proof of a generalized conjecture for Yukawa interactions,
working in arbitrary dimensions and at the local level of the heat kernel.
We will be able to perform a resummation of terms involving powers of the potential, together with its first and second derivatives.
Our results for the Yukawa interaction will then provide a shortcut to our second important achievement,
the proof of the generalized conjecture for SQED, again valid in arbitrary dimensions and for the diagonal of the heat kernel.
Finally, the previous results are discussed; several applications and possible future lines of studies are also mentioned.

Referring to the notation, we will use Planck units, for which $\hbar=1=c$.


\section{Resummation of the heat kernel for a Yukawa background}\label{sec:HK_Y}

For a Yukawa coupling, the heat kernel $K(x,x';\tau)$ defined in Eq.~\eqref{eq:effective_action_hk} satisfies the heat equation
\begin{align}\label{eq:HK_eq_yukawa}
 [\partial_\tau - \nabla_x^2 + V(x)] K_Y(x,x';\tau)&=0,
 \end{align}
 together with the following initial condition in the propertime $\tau$:
 \begin{align}\label{eq:HK_initial_condition}
 K_Y(x,x',0^+)= \delta(x-x').
\end{align}
These equations are of course analogous to those satisfied by a Schrödinger propagator in imaginary time and the results we are going to obtain could thus be employed in a first quantization.
Nevertheless, we will keep in mind the picture that $V$ in Eq.~\eqref{eq:HK_eq_yukawa} is to be regarded as the background value of the field $H$, which interacts with $\phi$ through the Yukawa coupling.
Note that we are employing a covariant notation even if we will be working in flat space.

As already stated above, we are interested in the resummation of contributions involving $V$, as well as those containing first and second derivatives of it, in a precise way that we are going to state shortly. First, let us recall some well-known facts. In the standard approach, when the potential includes a mass term
\begin{equation}\label{mass}
V(x) = M^2 + \varphi^2(x)\, ,
\end{equation}
the heat kernel can be written using the SDW Ansatz
\begin{align}\label{eq:HK_Y_mass}
 \begin{split}
  K(x,x';\tau)=& \frac{1}{(4\pi)^{\dime/2}} e^{-\tau M^2-\frac{\sigma(x,x')}{2\tau}}\widetilde \Omega(x,x';\tau),
  \end{split}
  \end{align}
where $\sigma(x,x')=(x-x')^2/2$ denotes Synge's function~\cite{DeWitt:2003}. The function $\widetilde\Omega_Y$ will in general depend on the background field $\varphi(x)$ and its derivatives. As noted some time ago in Refs.~\cite{Guven:1986gi, Paz:1988mt}, the powers of the background field can be resummed replacing $M^2$ by $V(x)$ in the exponent of the above equation\footnote{The replacement of $M^2$ by  $V(x')$ or by $(V(x)+V(x'))/2$ also works.}. After this replacement, $\widetilde \Omega_Y$ will have a small propertime expansion with coefficients whose coincidence limit will depend only on derivatives of the potential. Incidentally, the resummation of the Ricci scalar mentioned before can be achieved, in the case of spin-0 fields in curved spacetimes, by the replacement
$M^2\to M^2 +(\xi- 1/6)R$, see Refs.~\cite{Parker:1984dj, Jack:1985mw}.

The resummation of terms containing first and second derivatives of the potential is a more complex task.  It is natural to use as inspiration the simplest possible case for which these contributions do not vanish, i.e. the case of a harmonic potential. Considering the Fourier transform of the integrand in Eq.~(2.18) of Ref.~\cite{Brown:1975bc}, together with the definitions in Eq.~(2.30) of the same reference,\footnote{Note that, contrary to the Lorentzian approach in Ref.~\cite{Brown:1975bc}, here we are working in Euclidean space.} we are led to the Ansatz
\begin{align}\label{eq:HK_Y_ansatz}
 \begin{split}
  &K_Y(x,x';\tau)
  \\
  &=:\frac{1}{(4\pi)^{d/2}}\frac{e^{-\tau V(x')-\frac{1}{4}\tilde\sigma^\mu A^{-1}_{\mu\nu}(x';\tau)\tilde\sigma^{\nu} -C(x';\tau)}}{ \det^{1/2}\left(\tau^{-1} A(x'; \tau) \right)}\Omega_Y(x,x';\tau),
  \end{split}
  \end{align}
where we have defined the following quantities:
\begin{align}\label{eq:HK_Y_ansatz_dic}
\begin{split}
  \tilde\sigma_\mu(x,x'):&= \nabla_{\mu}\sigma(x,x')+ B_\mu(x';\tau),
  \\
  A_{\mu\nu}(x;\tau):&=\left[ \frac{1}{\gamma}\tanh(\gamma \tau)\right]_{\mu\nu},
  \\
  B_{\mu}(x;\tau):&= 2 \nabla^{\nu}V \left[\gamma^{-2}\left( 1-\sech(\gamma \tau)\right)\right]_{\nu\mu},
  \\
  C(x,\tau):&=  \nabla^{\mu}V \left[-\tau \gamma^{-2} +\gamma^{-3}\tanh(\gamma \tau)\right]_{\mu\nu} \nabla^{\nu}V
  \\
  &\hu+\tfrac{1}{2}\left[\log\bigl(\cosh(\gamma \tau)\bigr)\right]^{\mu}{}_{\mu},
  \\
  \gamma^2_{\mu\nu}:&=2\nabla_{\mu\nu}V.
 \end{split}
\end{align}
As a short-hand notation, $\nabla^{\alpha_1\cdots\alpha_n} X:= \nabla^{\alpha_1}\cdots\nabla^{\alpha_n} X$ is intended, which is symmetric in the present case since we are working in flat space, and we define $(\mathfrak{M}^n)^\alpha{}_{\beta}:=\mathfrak{M}^{\alpha}{}_{\mu_1} \mathfrak{M}^{\mu_1}{}_{\mu_2}\cdots \mathfrak{M}^{\mu_{n-1}}{}_{\beta}$ for the n-th power of an arbitrary matrix $\mathfrak{M}^{\alpha}{}_{\beta}$, which is understood to be extended for arbitrary functions through a series expansion (indices are lowered and raised with the flat metric $\delta_{\mu\nu}$). Furthermore, we will denote quantities evaluated in $x'$ with a prime,  for example, $(\gamma')^2_{\mu\nu}:=2\nabla_{\mu\nu}V(x')$. Finally, to avoid ambiguities, any time it is not clear from the context, we will explicitly provide the point at which the quantities are evaluated.

Several comments are in order with respect to Eq.~\eqref{eq:HK_Y_ansatz}. On the one hand, when the third- and higher-order derivatives of the potential vanish, $\Omega_Y(x,x';\tau)=\tau^{-d/2}$ is the exact solution for the heat kernel~\cite{Brown:1975bc}.
On the other hand, the Ansatz clearly goes beyond the SDW proposal, as one immediately realises that the exponential is already resumming contributions that would involve an infinite number of terms in the usual GSDW coefficients, e.g. the $e^{-\tau V'}$ term.
More importantly, its intrinsic structure is essentially different from the usual $e^{-\sigma/(2\tau)}$ prefactor,
which is independent of the background potential and whose dependence with $\tau$ is rather trivial. This will prove crucial in our proof below and it shows that our resummation is not just an exponentiation of GSDW coefficients, as considered in \cite{Barvinsky:2002uf}.

Replacing the Ansatz~\eqref{eq:HK_Y_ansatz} in the heat kernel equation, we get
\begin{align}\label{eq:HK_Y_prerecurrence}
 \begin{split}
&\Bigg\lbrace - \nabla^2+  \Big(\gamma' \coth(\gamma' \tau)-\frac{1}{\tau}\Big){}_{\alpha \beta } \nabla^\alpha\sigma(x,x') \nabla^{\beta }
\\
&\hu+ 2 {\nabla}^{\beta }V(x')  \left(\frac{\tanh(\gamma' \tau/2)}{\gamma'}\right)_{\beta \alpha } \nabla^{\alpha }
\\
&\hspace{1cm}+ \mathfrak{S}(x,x';\tau)\Bigg\rbrace \Omega_Y (x,x';\tau)
\\
&=\left\lbrace -\frac{\dime}{2\tau} -\partial_\tau -\frac{1}{\tau} \nabla^\alpha\sigma(x,x') \nabla_{\alpha } \right\rbrace  \Omega_Y (x,x';\tau),
\end{split}
\end{align}
where we have also defined the effective potential
\begin{align}\label{eq:frakS}
\begin{split}
\mathfrak{S}(x,x';\tau):&=
  V(x) -  V(x')
-  \nabla^\alpha\sigma(x,x') {\nabla}_{\alpha }V(x')
\\
&\hspace{1.0cm}-\frac{1}{4}  \nabla^{\alpha}\sigma(x,x')\nabla^\beta \sigma(x,x') (\gamma')^2_{\alpha \beta }.
\end{split}
\end{align}
Not so unexpectedly, expanding Eq.~\eqref{eq:HK_Y_prerecurrence} in $\tau$ one obtains only even powers of $\gamma$,
which enforces the GSDW coefficients to depend only on natural powers of the derivatives of the potential.

At this point it is opportune to state our claim more precisely: we will prove that, after expanding $\Omega_Y$ in powers of the propertime,
\begin{align}\label{eq:HK_Y_expansion}
    \Omega_Y =:\sum_{j=0}^\infty a_j(x,x') \tau^{j-d/2},
\end{align}
none of the coincidence limits
\begin{align}
   [a_j]:=[a_j(x,x')] := \lim_{x'\to x} a_j(x,x') , \quad j\geq 0,
\end{align}
will depend on the geometric invariants contained in the set
\begin{align}\label{eq:invariants_Y}
    \mathcal{K}_Y=\lbrace V, \; \delta^{\alpha\beta}\gamma^{j}_{\alpha\beta}, \;\nabla^{\alpha}V\gamma^{j}_{\alpha\beta}\nabla^\beta V, \quad j\geq 0\rbrace.
\end{align}
To simplify the discussion, we will refer to the elements of $\mathcal{K}_Y$ as \emph{chains}.
Of course, our assertion does not preclude the appearance of the first or second derivatives of the potential in an invariant in which a higher derivative is present.
Notice also that the invariant $(\nabla V)^2$ is contained as the limiting $j=0$ case in the set $\mathcal{K}_Y$.

The proof goes as follows. First, using Eq.~\eqref{eq:HK_Y_expansion}, we expand the expression~\eqref{eq:HK_Y_prerecurrence} in powers of the propertime $\tau$.
Equating the coefficients with equal powers of $\tau$, we obtain a relation between $a_{j+1}$ and the (derivatives of the) previous coefficients,
\begin{align}
\begin{split}\label{eq:HK_Y_recurrence}
&-\left( j+1+\nabla_\alpha\sigma \nabla^\alpha \right) a_{j+1}(x,x')
= (-\nabla^2 +\mathfrak{S}) a_{j}(x,x')
\\
&\hspace{0.5cm}+ \sum_{n=1}^{\lfloor j/2 \rfloor} \frac{B_{2n}}{(2n)!}
\Big( 4(2^{2n}-1) \nabla^\alpha V' \left(\gamma'\right)_{\alpha\beta}^{2(n-1)}
\\
&\hspace{1.7cm}+2^{2n} \nabla^\alpha\sigma \left(\gamma'\right)^{2n}_{\alpha\beta}\Big)
\nabla^\beta a_{j+1-2n}(x,x'),
\end{split}
\end{align}
where $B_n$ is the $n$th Bernoulli number and $\lfloor \cdot \rfloor$ is the floor function.
The expression~\eqref{eq:HK_Y_recurrence} enables one to recursively compute the coincidence limit of the heat kernel coefficients $a_j$,
using the fact that the initial condition~\eqref{eq:HK_initial_condition} implies
\begin{equation}\label{eq:a0}
a_0(x,x')=1.
\end{equation}
In performing the
calculation, one also needs the derivatives of the previous coefficients,
for which a recursive formula can be obtained differentiating Eq.~\eqref{eq:HK_Y_recurrence}.
In this way, an evidently well-defined ordering is created, which starts as $[a_0]$, $[a_1]$, $[\nabla_\mu a_1]$, $[\nabla_{\mu\nu}a_1]$, $[a_2]$, $\cdots$. It is worth noting that the calculation proceeds in a way analogous to the usual SDW approach; the difference resides in the appearance of the $\mathfrak{S}$ term and the sum over $n$ in Eq.~\eqref{eq:HK_Y_recurrence},
involving derivatives of possibly all the previous coefficients.

Coming back to the proof, after taking the coincidence limit in Eq.~\eqref{eq:HK_Y_recurrence} one notices that none of the chains appears explicitly in the recurrence. Moreover, one can also take derivatives in Eq.~\eqref{eq:HK_Y_prerecurrence} and express $\nabla^{\alpha_1\cdots\alpha_n} a_j$ in terms of derivatives of the previous coefficients; after taking the coincidence limit, no chain will explicitly appear.
Indeed, the only possible source of these invariants are the contributions produced by $\mathfrak{S}$; the structure \eqref{eq:frakS} of $\mathfrak{S}$, however, is such that not only its coincidence limit but also those of the first and second derivative vanish, $[\mathfrak{S}]=[\nabla_\mu\mathfrak{S}]=[\nabla_{\mu\nu}\mathfrak{S}] =0$. Only third and higher-order derivatives of $\mathfrak{S}$ have nontrivial coincidence limits, containing third or higher-order derivatives of $V$. This is a crucial advantage of the Ansatz~\eqref{eq:HK_Y_ansatz} in comparison with the usual SDW expansion.

Still, chains could arise implicitly generated by the structure of the (derivatives of the) previous coefficients. By inspection of Eq.~\eqref{eq:HK_Y_recurrence}, we see that this is possible only if derivatives of previous coefficients or contractions thereof contain ``half-chains'' of the invariants in $\mathcal{K}_Y$,
to wit $ \gamma^{2l}_{\alpha\beta}$ or $\gamma^{2l}_{\alpha\beta}\nabla^\beta V$ with $l= 0,1,\cdots$.
By induction, taking as ordering the one dictated by the recursion in Eq.~\eqref{eq:HK_Y_recurrence}, one can prove that this is not going to be the case.

In effect, for the first element in the order, i.e. $[a_0]$, this is true. Afterwards, taking the derivatives $\nabla^{\alpha_1\cdots\alpha_n}$ on both sides of Eq.~\eqref{eq:HK_Y_recurrence} and the coincidence limit, on the LHS we will obtain the quantity to be computed at this order, namely $\nabla^{\alpha_1\cdots\alpha_n} a_{j+1}$, times a numerical factor. On the first term of the RHS, we will obtain $\nabla^{\mu}{}_{\mu}{}^{\alpha_1\cdots\alpha_n}a_{j}$, which is a previous element in the inductive order, and therefore upon contraction can not contribute with a half-chain because of the inductive assumption. The remaining contribution in the first line, i.e. the term proportional to $\mathfrak{S}$, will not contribute because of the inductive assumption and the comments in the previous paragraphs.

After derivatives and the coincidence limit are taken, the sum over $n$ in Eq.~\eqref{eq:HK_Y_recurrence} will contain contributions which are of the schematic form $(\gamma^{2m})^{\alpha_1}{}_{\beta} \nabla^{\beta\alpha_2\cdots\alpha_n}a_{l}$ for some $m,l,n\geq 0$. If this is going to contain a half-chain, then the half-chain could involve the $(\gamma^{2m})^{\alpha_1}{}_{\beta}$ prefactor or not. If it does not, then the half-chain is completely contained in $\nabla^{\beta\alpha_2\cdots\alpha_n}a_{l}$, which conflicts with the inductive assumption. If it does, then we have yet a contradiction, since to form a half-chain from $(\gamma^{2m})^{\alpha_1}{}_{\beta}$ we forcedly need another half-chain.

This finishes the perturbative proof of the resummation for the Yukawa case.
The validity of our Ansatz in Eq.~\eqref{eq:HK_Y_ansatz} can be checked computing the first coefficients $a_j$; after expanding the whole expression~\eqref{eq:HK_Y_ansatz} in the propertime and taking the coincidence limit, one should obtain the usual GSDW coefficients. Our results are in agreement with Ref.~\cite{Vassilevich:2003xt} up to the third-order coefficients (the highest order it provides). Using the results in Ref.~\cite{vandeVen:1997pf}, we can compute the coefficients up to the fifth order, which are also in agreement with our resummed expressions (see the comments in~\cite{Franchino-Vinas:2024wof}).
Note that no integration by parts is used in these comparisons.


\section{Resummation of the heat kernel for an electromagnetic background}\label{sec:HK_EM}
  Consider now the heat kernel of a Laplace-type operator as in Eq.~\eqref{eq:op}, in this case involving just an electromagnetic field.
 Upon expansion of the covariant derivative, the heat kernel's equation  can be recast in this case as
 \begin{align}\label{eq:HK_eq_EM}
 [\partial_\tau - \nabla^2 +2A_\mu(x) \nabla^\mu + V_{EM}(x)] K_{EM}(x,x';\tau)&=0,
 \end{align}
 where the effective scalar potential is given by
 \begin{align}\label{eq:VforEM}
 V_{EM}(x):&= \nabla_\mu A^\mu-A_\mu A^\mu.
 \end{align}
We choose the Fock--Schwinger gauge for the electromagnetic potential, i.e.
\begin{align}
 (x-x')^\mu A_\mu(x)=0,
\end{align}
which will prove convenient for the following reasons.
 Using the results in Ref.~\cite{Pascual:1984zb}, we can show that the Fock--Schwinger condition is equivalent to the following series expansion of the electromagnetic potential around the point $x'$:
\begin{align}
 \begin{split}\label{eq:A_fock_series}
A_\mu(x'+x) &=
\sum_{n=0}^{\infty} \frac{1}{n! (n+2)} x^{\rho} x^{\mu_1}\cdots x^{\mu_n} \nabla_{\mu_1\cdots \mu_n}F_{\rho\mu}(x').
 \end{split}
\end{align}
This formula provides an expansion of the potential in terms of the field strength; in particular, the derivatives of $A_\mu$ in the coincidence limit take the simple form
\begin{align}\label{eq:coincidenceA_Field}
[A^{\nu}{}_{;\mu_1\cdots\mu_n}(x)]=\frac{n}{n+1} F_{(\mu_1}{}^{\nu}{}_{;\mu_2\cdots \mu_n)}(x),
\end{align}
where idempotent symmetrization has been denoted in the RHS of Eq.~\eqref{eq:coincidenceA_Field} by writing the indices between parentheses.

Let us state our claim for this case: we will show that there exists a resummation such that no geometric invariant contained in \begin{align}\label{eq:invariants_EM}
    \mathcal{K}_{EM}=\lbrace (F^n)^{\mu}{}_{\mu}, n\geq 0\rbrace
\end{align}
appears in the coefficients $[a_j]$ of the heat kernel.
Since
\begin{align}\label{eq:gamma_EM}
    \gamma^2_{\mu\nu}=(F^2)_{\mu\nu}/4+(\text{\rm derivatives of $F$}),
\end{align}
resumming the contributions of $V_{EM}$ to the heat kernel [including up to its second derivatives, as done in Eq.~\eqref{eq:HK_Y_ansatz}], we will be resumming the desired contributions.
In effect, although there could arise contributions from the term in Eq.~\eqref{eq:HK_eq_EM} which is linear in $A_\mu$, we will now show that this is not the case.

Consider thus the Ansatz in Eq.~\eqref{eq:HK_Y_ansatz}, substituting $V$ with $V_{EM}$. Inserting this Ansatz in Eq.~\eqref{eq:HK_eq_EM}, we obtain an expression which contains all the terms in Eq.~\eqref{eq:HK_Y_prerecurrence}; in addition, it also possesses the following contributions, which come from the  term in Eq.~\eqref{eq:HK_eq_EM} proportional to $A_\mu$ (we are neglecting the unnecessary overall factors):
\begin{align}\label{eq:HK_eq_EM_linear}
\begin{split}
&{2A_\mu(x) \nabla^\mu  K_{EM}(x,x';\tau)}
\\
&\sim \Bigg\lbrace  2A^{\alpha }(x) \nabla_{\alpha } - A^{\alpha }(x) \Big(\gamma' \coth(\gamma' \tau)\Big)_{\alpha\beta} \nabla^{\beta}\sigma(x,x')\\
&\hspace{0.5cm}
-2A^{\alpha }(x) \left(\frac{1}{\gamma'}\tanh(\tfrac{1}{2} \gamma' \tau)\right)_{\alpha\beta}{\nabla}^{\beta}V_{EM}(x')
\Bigg\rbrace
\\
&\hspace{0.8cm}\times\Omega_{EM} (x,x';\tau).
\end{split}
\end{align}
This expression can be analyzed term by term as has been done for the Yukawa potential.
Beginning with the third term in the RHS of Eq.~\eqref{eq:HK_eq_EM_linear},
a simple computation shows that, because of $\nabla^\beta V_{EM}$, it contributes with an invariant containing derivatives of the field strength,
no matter whether we consider Eq.~\eqref{eq:HK_eq_EM_linear} or its derivatives.

To deal with the second term in the RHS of Eq.~\eqref{eq:HK_eq_EM_linear}, it is enough to consider just the contributions in $\gamma_{\alpha\beta}$  which contain no derivatives of the field strength; taking an arbitrary number of derivatives we find\footnote{The subindex $(n)$ labels to which term in Eq.~\eqref{eq:HK_eq_EM_linear} we are referring.}
\begin{align}
 \begin{split}
&\nabla^{\alpha_1\cdots\alpha_n}\Big( A_\mu(x) \nabla^\mu  K_{EM}\Big)_{(2)}
 \\
 &\hspace{0.5cm}\to F^{\beta\alpha_1} (F^{2n})_{\beta}{}^{\alpha_2} \nabla^{\alpha_3\cdots\alpha_n}\Omega_{EM}, \quad n=0,1\cdots.
 \end{split}
\end{align}
A chain could arise explicitly contracting with $\delta_{\alpha_1\alpha_2}$. However, this term is antisymmetric in $(\alpha_1,\alpha_2)$,
given that it contains an odd power of the antisymmetric tensor $F$; given that the derivatives were symmetric in the same indices, it must vanish. If instead the chain arises implicitly from $\nabla^{\alpha_3\cdots\alpha_n}\Omega_{EM}$, then it would be in contradiction with the inductive assumption.

The last possible source of chains is the first term in the RHS of Eq.~\eqref{eq:HK_eq_EM_linear}.
In this case we appeal once more to an induction proof. Following the ideas in the previous section, assume that half-chains of the form $(F^n)_{\alpha\beta}$ do not appear in contractions of the derivatives of the coefficients up to the point in which we compute (derivatives of) the coefficient $a_j$. Then, (the derivatives) of the first term in the RHS of Eq.~\eqref{eq:HK_eq_EM_linear} could contribute only in the form:
\begin{align}
    \nabla^{\alpha_1\cdots\alpha_n}\Big( A_\mu(x) \nabla^\mu  K_{EM}\Big)_{(1)} \to F_{\alpha}{}^{\alpha_1} \nabla^{\alpha\alpha_2\cdots\alpha_n}a_{j-1}.
\end{align}
But then the only way to form a half-chain  upon contractions is that they should be present in contractions of $\nabla^{\alpha\alpha_2\cdots\alpha_n}a_{j-1}^{(EM)}$ itself, which would be impossible by assumption.
Adding the fact that also in the electromagnetic case the initial condition~\eqref{eq:HK_initial_condition} forces $a_0(x,x')=1$, the proof is complete.

We have checked the validity of our results by computing the first heat kernel  coefficients.
Without employing integration by parts but using the Bianchi identity, our results agree with Ref.~\cite{Vassilevich:2003xt} up to third order in the propertime, which is the highest order it provides. We can also compute the standard GSDW coefficients using the standard methods described in Ref.~\cite{vandeVen:1997pf}; up to the fifth order and taking into account the comments in Ref.~\cite{Franchino-Vinas:2024wof}, they agree with our results.
Alternatively, one can use integration by parts to successfully compare with Refs.~\cite{Fliegner:1997rk, Navarro-Salas:2020oew}, again up to and including the fifth-order coefficient.
To successfully compare with Ref.~\cite{Navarro-Salas:2020oew},
it is important to note that the basis of invariants that they consider is overcomplete;
indeed, using the Bianchi identities we can prove that two of them are related with the remaining ones:
\begin{align}
F^{\kappa \lambda } F^{\mu \nu } \nabla_{\rho }F_{\lambda \nu } \nabla^{\rho }F_{\kappa \mu }
&= 2 F^{\kappa \lambda } F^{\mu \nu } \nabla_{\nu }F_{\lambda \rho } \nabla^{\rho }F_{\kappa \mu },
\\
\begin{split}
F^{\kappa \lambda } F^{\mu \nu } \nabla_{\nu }F_{\lambda \rho } \nabla^{\rho }F_{\kappa \mu }
&
\\
&\hspace{-3cm}=
-F^{\kappa \lambda } F^{\mu \nu } \nabla_{\lambda }F_{\nu \rho } \nabla_{\mu }F_{\kappa }{}^{\rho } + F^{\kappa \lambda } F^{\mu \nu } \nabla_{\mu }F_{\kappa }{}^{\rho } \nabla_{\nu }F_{\lambda \rho }.
\end{split}
\end{align}
Even after taking into account these equalities (or integration by parts) our result still does not agree with Ref.~\cite{Gusynin:1998bt}.
As far as we can see, the fact that they employ a derivative expansion should not be an obstacle and our results should coincide in the appropriate range of validity. Given our agreement with the other references, it seems probable that there could have been a numerical error in some intermediate steps of Ref.~\cite{Gusynin:1998bt} or a subtlety in the application of the Wordline Formalism has passed unnoticed.


\section{Discussion and Conclusions}\label{sec:conclusions}

Summarizing, we have proved that the diagonal of the heat kernel, both for a Yukawa and an electromagnetic background, satisfies the resummation formula
\begin{align}\label{eq:HK_diagonal}
\begin{split}
&K(x,x;\tau)
\\
&=
\frac{e^{-\tau V +\nabla^{\alpha }V \left[ \gamma^{-3} \left({\gamma \tau - 2 \tanh(\tfrac{1}{2} \gamma \tau)}\right)\right]_{\alpha \beta } \nabla^{\beta }V }}
{(4\pi)^{\dime/2} \;{\det} ^{1/2}\big((\gamma \tau)^{-1} \sinh(\gamma \tau)\big) }  \Omega(x,x;\tau),
\end{split}
\end{align}
where the explicit form of $V$, $\gamma$ and $\Omega$ depends on the case under consideration.
The prefactor in Eq.~\eqref{eq:HK_diagonal} should be understood as a resummation of an infinite number of invariants;
a nontrivial cancellation of singular terms in the exponent is responsible for its regular character for small $\tau$.

In the Yukawa case, the prefactor comprises all the invariants that can be built using solely the potential and its first or second derivatives, cf. Eq.~\eqref{eq:invariants_Y}.
If there exist physical reasons to dismiss second derivatives, one can take the smooth limit for small $\gamma$  in Eq.~\eqref{eq:HK_diagonal} to obtain
\begin{align}\label{eq:HK_diagonal_first_order}
K(x,x;\tau)&=\frac{1}{(4\pi)^{\dime/2}}e^{-\tau V +\frac{\tau^3}{12}\nabla^{\alpha }V  \nabla_{\alpha }V } \Omega'(x,x;\tau),
\end{align}
where now $\Omega'(x,x;\tau)$ will depend in general on all the invariants built from $\nabla_{\mu\nu}V$. In this case, the form of the prefactor
could have been guessed from the expression of $a_3$ in the simpler GSDW expansion.
It is also important to bear in mind that the invariant $(\nabla V)^2$ is always present in Eq.~\eqref{eq:HK_diagonal},
in the sense that it appears upon a small-propertime expansion.

In the electromagnetic case, all the chains $(F^n)^{\mu}{}_{\mu}$, $n=1,2,\cdots$ have been resummed.
An immediate consequence is the proof of the conjecture stated in Ref.~\cite{Navarro-Salas:2020oew},
which, using our notation, reads as follows:
the coefficients in the small-propertime expansion of the function $\Omega_{EM}$ appearing in our Eq.~\eqref{eq:HK_eq_EM_linear} do not depend on the invariants $\mathcal{F}:=F^{\mu\nu}F_{\mu\nu}$ and $\mathcal{G}:=\tilde{F}^{\mu\nu}F_{\mu\nu}$.
In effect, this is true since we have proved that all the invariants in $\Omega_{EM}$ contain at least one derivative of the field strength,
while $\mathcal{F}$ and $\mathcal{G}$ contain none.
In four dimensions, one can obtain the usual Euler--Heisenberg Lagrangian using the fact that invariants built only from the field strength can be expressed in terms of just $\mathcal{F}$ and $\mathcal{G}$ (see Ref.~\cite{Landau:Vol2}).

Coming back to generalities, it is important to emphasize that the result in Eq.~\eqref{eq:HK_diagonal} is local,
implying that it can be used to compute propagators in the coincidence limit or smeared traces $\operatorname{Tr}\big(f(x) K(x,x';\tau)\big)$, with $f(x)$ a sufficiently convergent function at infinity.
Furthermore, it drastically reduces the number of invariants involved in an expansion of $\Omega$ in the propertime.
For instance, consider the Yukawa case: the third order contribution contains four terms in the usual SDW expansion~\cite{Vassilevich:2003xt}, which reduces to just one in our setup.
This reduction is more substantial for higher-order coefficients,
for which it is known that the number of contributing invariants rapidly increases; for instance, $[a_5]$ reduces from sixteen  to five terms.

For small propertime $\tau$, the leading behaviour of the expression~\eqref{eq:HK_diagonal} is as usual independent of the background potential and dictated by the dimension of the manifold, cf. expression~\eqref{eq:HK_Y_expansion}. Note that, as already mentioned,  the prefactor in Eq.~\eqref{eq:HK_diagonal} is smooth in the limit when $\gamma$ (or $\gamma \tau$) tends to zero.
Additionally, no singularities arise as long as $\gamma$ has positive eigenvalues (which will be true if the Hessian of $V$ is positive; this is rather intuitive if we think of the spectral decomposition of the heat kernel).
Instead, if the Hessian of $V$ possesses at least one negative eigenvalue $\lambda_-$, the determinant in Eq.~\eqref{eq:HK_diagonal} develops an infinite number of poles in the propertime determined by the condition $\sin(\sqrt{|\lambda_-|}\tau)=0$.

As we will shortly see, this is the quintessence of the Schwinger effect. To this end, compute the effective action as discussed in Sec.~\ref{sec:intro};
in the Euclidean setup we have
\begin{align}
\Gamma_{\rm E}=- \int \dxd[x][\dime] \int_0^\infty \frac{\dx[\tau]}{\tau}  K(x,x;\tau).
\end{align}
If we want to analyze the Minkowskian situation, it is customarily to perform a Wick rotation,
which basically consists in substituting $t\to \mathi t$ and in simultaneously introducing a factor $\mathi$ for every single zeroth-component in all the involved tensors.
This could give rise to negative eigenvalues in some region of the spacetime, thus creating instabilities in the Hessian of V, even if it was positive definite in Euclidean space.
Hence, we have
\begin{align}
\Gamma_{\rm M}\sim \int \dxd[x][\dime] \int_0^\infty \frac{\dx[\tau]}{\tau}  {{\det} ^{-1/2}\big[(\tilde\gamma_\epsilon \tau)^{-1} \sinh(\tilde\gamma_\epsilon \tau)\big] } \times \text{``reg.''},
\end{align}
where ``reg.'' denote terms that we assume to be regular and nonvanishing at the poles of the determinant where $\tau\neq0$ (and whose real nature can not be unveiled unless we have a resummed/exact result), while $\tilde\gamma_\epsilon$ is related to the Wick-rotated second derivative of the potential, keeping a small parameter $\epsilon$ to circumvent the poles (it is natural to simply understand $\epsilon$ as stopping the Wick rotation just before reaching the axis of imaginary time).
As a consequence, $\Gamma_M$ acquires an imaginary contribution, which signals an instability of the vacuum and is associated with the creation of pairs; in effect, in the limit of a weak pair creation, the total pair creation probability reads
\begin{align}
P =2 \operatorname{Im} \Gamma_M.
\end{align}
In the electromagnetic case, since essentially $\gamma_{\mu\nu}\sim (F)^2_{\mu\nu}$, cf. Eq.~\eqref{eq:gamma_EM}, this clearly happens when the electric field dominates over the magnetic field: this is the well-known Schwinger effect.

However, the mechanism also gives rise to a scalar Schwinger effect,
i.e. one in which the background is scalar.
From our perspective, the unified origin of the scalar and electromagnetic Schwinger effect is patent.
A simple example within our scalar approach is provided by inflaton models in which the scalar field slowly rolls down  with a temporal profile that could be approximated as linear.
More complex models are also in the range of applicability of our results; for instance, the study of ultralight dark matter creation through a Higgs portal~\cite{Patt:2006fw, Piazza:2010ye}.
Another possibility, though technically inaccessible in the laboratories with the present technology,
is to consider pulses of Higgs particles, consisting of a large, coherent number thereof,
which would then create a suitable scalar background for pair creation.

Coming to future work, it would be beneficial to explore how the structure of the Worldline's Green function, as discussed in Ref.~\cite{Gusynin:1998bt}, precludes the formation of chains at higher orders in the derivative expansion.
In our case, the key ingredient to perform the proof was the use of the Ansatz~\eqref{eq:HK_Y_ansatz}, which filters into the perturbative computation through the effective potential $\mathfrak{S}$ in Eq.~\eqref{eq:frakS}.

Further developments could involve complementing the ideas in this article with numerical implementations of the Worldline Formalism, offering a semi-analytical (or semi-numerical) approach to obtain nonperturbative results;
by this, we mean that a large piece of information can be added analytically before applying the numerical methods, as in the renormalization process to obtain a two-point function~\cite{Franchino-Vinas:2019udt}.
Implementing the factorization that we have shown to be valid in this letter could be employed at least to reduce the computation time, similarly to what happens in the quantum mechanical situation presented in Ref.~\cite{Ahumada:2023iac}.

Extending the framework to Yang--Mills and Proca-like backgrounds seems possible under reasonable assumptions. Another promising idea regards higher-derivative operators: it is known that for these operators,
which appear in theories of supergravity and super Yang--Mills~\cite{Beccaria:2015ypa, Casarin:2019aqw, Casarin:2023ifl},
the small propertime asymptotics is intrinsically modified. It would be interesting to see how these modifications affect the possible resummations.

From an experimental perspective, our resummed version of the proper-time expansions paves the way for an enhanced analysis of several observationally accessible physical phenomena: on the one hand, given the local nature of the expansions, they could provide an alternative formulation to the locally-constant field approximation frequently employed in strong field QED for the computation of the pair creation rate~\cite{Baier:1988dp}, presenting a valuable prospect to the pressing need in the field of improved approximations~\cite{Salgado:2021uua, Ilderton:2018nws, DiPiazza:2017raw}. On the other hand, they also explicitly harbour electromagnetic nonlinearities and hence they can be straightforwardly associated with testable nonlinear QED processes, such as light-by-light scattering~\cite{ATLAS:2017fur,ATLAS:2019azn} or vacuum birefringence~\cite{Karbstein:2015xra}.


\section*{Acknowledgments}
SAF acknowledges the support from Helmholtz-Zen\-trum Dresden-Rossendorf.
SAF and UWH acknowledge the support of UNLP through the project 11/X748.
SAF and FDM acknowledge the support through the project PIP 112\-20200101426CO from Consejo Nacional de Investigaciones Cien\-tí\-ficas y Técnicas (CONICET).
The work of VV has been partially funded by Next Generation EU through
the pro\-ject ``Geometrical and Topological effects on Quantum Matter (Ge\-TOnQuaM)''.
The research activities of CGP and VV have been carried out in the framework of the INFN Research Project QGSKY. The Authors extend their appreciation to the Italian National Group of Mathematical Physics (GNFM, INdAM) for its support.
The authors would like to acknowledge the contribution of the COST Action CA18108 ``Quantum gravity phenomenology in the multi-messenger approach''.


\bibliographystyle{elsarticle-num}
\bibliography{bibliografia}

\end{document}